# A-GPS Assisted Wi-Fi Access Point Discovery on Mobile Devices for Energy Saving

Feng Xia, Wei Zhang, Fangwei Ding, Ruonan Hao
School of Software, Dalian University of Technology, Dalian 116620, China
Email: f.xia@ieee.org

*Abstract*—Mobile devices have been shipped with multiple wireless network interfaces in order to meet their diverse communication and networking demands. In this paper, we propose an A-GPS assisted scheme that discovers the nearest Wi-Fi network access points (APs) by using user's location information. This allows the user to switch to the Wi-Fi interface in an intelligent manner when she/he arrives at the nearest Wi-Fi network AP. Therefore, it avoids the long periods in idle state and greatly reduces the number of unnecessary Wi-Fi scans on the mobile device. The experimental results demonstrate that our scheme effectively saves energy for mobile devices integrated with Wi-Fi and cellular interfaces.

## I. INTRODUCTION

Many wireless network technologies, such as bluetooth, 802.11 Wi-Fi, and GPRS, have become popular in the market. According to [1, 2, 3, 4], Wi-Fi and GPRS have widely different characteristics in terms of power, range and bandwidth. The GPRS has low transfer power efficiency, wide range, and low bandwidth. On the contrary, the 802.11 Wi-Fi has high transfer power efficiency, narrow range, and high bandwidth. However, it has high power consumption in idle state and brings a high overhead when scanning for new networks.

As for mobile devices, such as smartphones, the lifetime of the battery is more and more important to the users. For example, if the Wi-Fi is used all the time, the mobile device can work for only several hours before its energy become exhausted. Thus, one should leverage the complementary characteristics of these radios to meet the user bandwidth needs and minimize the power consumption.

In this paper, we address this challenge by presenting a simple scheme that finds the nearest Wi-Fi network access point (AP) by using the user's location information. Thus, users switch to Wi-Fi network when they arrive at it and greatly reduce the numbers of unnecessary Wi-Fi scans on the mobile devices. Therefore, we can save much energy for the users.

Currently, many location sensing modalities, such as GPS and GSM, have been developed for mobile devices [5, 6]. As described in [7], GPS and GSM have widely different characteristics in relation to power and accuracy (see Table I). GPS has high accuracy, but it is power hungry. While GSM is power efficient, it is highly inaccurate. However, Assisted GPS (A-GPS for short) is same as GPS in terms of accuracy and with less power. Considering energy-accuracy tradeoff, we adopt A-GPS to get user's location information.

TABLE I. VARIOUS SENSOR CHARACTERISTICS

| Sensor | Average power consumption (mW) | Approximate accuracy (m) |
|---|---|---|
| GPS | 400 | 10 |
| GSM | 60 | 400 |

We make the following contributions. First, we advocate the use of Wi-Fi network for users who need a high bandwidth. Second, we propose an A-GPS assisted scheme that greatly reduces the numbers of unnecessary Wi-Fi scans on the mobile devices and our experiment demonstrates it's effective. Finally, we present cloud sharing to discovery Wi-Fi networks and offload computations to the cloud computing platform. Thus we can save more energy.

The rest of the paper is organized as follows. Section 2 describes the related work in reducing energy consumption for mobile devices. Section 3 deals with the problem of Wi-Fi discovery and also addresses collaborative discovery. The key components and operation workflow of our proposed system are discussed in Section 4. Three switching schemes and two non-switching schemes are described in Section 5 and evaluation of the different schemes is done in Section 6. Section 7 discusses extensions to our system along with future work, and we conclude the paper in Section 8.

## II. BACKGROUND

With the development of the wireless network and the mobile devices, more and more people will use smartphones not only for calling, but also for entertainment. More and more applications are developed, such as watching films on the web, downloading games from Internet. People like do this no matter anywhere. Such as they are walking on the street, or they are chatting in the Starbucks. But the GPRS's bandwidth is too low to do this. While the Wi-Fi has enough bandwidth to satisfy the users, it costs too much energy for users, and it's not available everywhere. Thus, how to make use of their advantages is a key problem for the users. This problem greatly limits the development of the smartphones' applications. How

This work was partially supported by the National Natural Science Foundation of China under Grant No. 60903153, the Fundamental Research Funds for the Central Universities (DUT10ZD110), and the SRF for ROCS, SEM, China.

to address this challenge will be conducted a detailed discussion in this paper.

It is commonly considered that Wi-Fi has rather high energy consumption. And the measurement shows that Wi-Fi requires 1.4260W [8] to scan and requires 0.890W [3] for active transmission. But it's only need 0.256W in the idle state. In [9] Zhang et al. estimate the idle state to save the energy consumption of the Wi-Fi radio interface. This will save the energy consumption. However, it's hard to precisely estimate the idle state. And Shih et al. [10] suggest that completely power off the radio interface and turn on the radio only when there is on-going traffic. But this will cause the user complete lose connection to the wireless LAN. In [11] Rozner et al. propose NAPman which leverages AP virtualization and a new energy-aware scheduling algorithm to minimize Wi-Fi devices power consumption.

Bluetooth's energy consumption is very low, so in [3], Pering et al. switch between bluetooth and Wi-Fi to save energy consumption. But, this scheme needs to modify the infrastructure. While in [8], Wu et al. use the GSM signal to get the location information to estimate the Wi-Fi network AP. The GSM's energy consumption is very small, but its positioning accuracy is very bad. In [12] Zhou et al. utilize ZigBee radios to identify the existence of Wi-Fi networks through unique interference signatures generated by Wi-Fi beacons. Even though they detect Wi-Fi network APs with high accuracy, short delay and little computation overhead, they incurs extra fees to buy or employ ZigBee cards.

There is an idea of using a separate low-powered radio to wake up a high-powered radio. Like [10] Wake-on-Wireless proposes the use of a low power radio that serves as a wake-up channel for a Wi-Fi radio. But it needs significant modifications to existing mobile devices. While On-Demand-Paging [13] builds on this idea to use the widely available bluetooth radio as the low-powered channel, it also requires substantial infrastructure support in the form of specialized APs that have both Wi-Fi and bluetooth interfaces. Cell2Notify [14] uses the cellular interface to wake up the Wi-Fi interfaces on an incoming VOIP call using specialized servers. Blue-Fi [4] doesn't need any modification to the existing infrastructure. But many mobile devices with bluetooth radio change their positions and bluetooth has a much lower range compared to other network technologies [4], this will result in that mobile devices with bluetooth can't be used for our problem.

## III. WI-FI DISCOVERY

In a nutshell, we discover the Wi-Fi network by periodically scan its signal. When a mobile device discovers a Wi-Fi network, it logs this Wi-Fi network in a log $L$, locally. The log entries are of the form (*Timestamp*, {*Wi-Fi network*}, *Location*, *Range*). Wi-Fi network is identified by its SSID/BSSID and location is identified by Wi-Fi network AP's geographical position. Range represents how far the Wi-Fi network signals reach. Given a user's location, we just check the log whether the user is in the range of a Wi-Fi network: if yes, then we consider the user can connect to the Wi-Fi network.

### A. Collaborative Discovery

In this section, we explores the scenarios and benefits if mobile devices were to collaborative and share information about Wi-Fi networks. For example, such sharing is beneficial when a user go to a new place without prior context. Basically, sharing speeds up the learning process. We present three sharing approaches which we have not implemented yet as follows.

*1) Peer-to-peer Sharing:* Devices query each other for the information of Wi-Fi networks over bluetooth. Any device which can be connected through bluetooth can respond with its local log $L$. Ideally, when a user arrive at a new place, he or she can get the information of Wi-Fi networks around. However, due to the bluetooth devices' short range and mobility, peer-to-peer sharing over bluetooth is not a good fit for our problem, especially in outdoors.

*2) Global Sharing:* Global sharing is facilitated through a central service. Devices, when using the Wi-Fi network, periodically upload entries − (*Timestamp*, {*Wi-Fi network*, {*Characteristics*}}) − to the centralized server. Characteristics include the Wi-Fi network's location, range and authentication information if required. Servers index these entries by the Wi-Fi network AP's location for efficient retrieval. Any device that wants to know about its Wi-Fi networks around can query the central server by supplying its geographical position. The server responds by matching the geographical position in its database and returns the Wi-Fi networks along with their characteristics. Devices communicate with the server through the cellular interfaces. The matching mechanism simply returns the nearest Wi-Fi network AP according to the user's location information. Thus, we can further save energy for users by offloading matching computations to the centralized server.

*3) Cloud Sharing:* A similar choice to global sharing is cloud sharing facilitated by a cloud computing platform. Like global sharing, any device, which can connect to the network, periodically uploads entries − (*Timestamp*, {*Wi-Fi network*, {*Characteristics*}}) − to the cloud computing platform. Characteristics are same as global sharing's. Cloud computing platform also indexes these entries by the Wi-Fi network AP's location for efficient retrieval. The matching mechanism is simply as same as the global sharing. But, due to the more powerful computing capability of cloud computing platform, it is faster and save more energy for users than global sharing.

The ability to offload computations makes the global sharing and cloud sharing options potentially more useful than peer-to-peer sharing. However, peer-to-peer sharing does not need any infrastructure support or money and hence is more readily deployable.

## IV. SYSTEM OVERVIEW

This section shows the key components and typical, operation workflow of the proposed system, see Fig. 1.

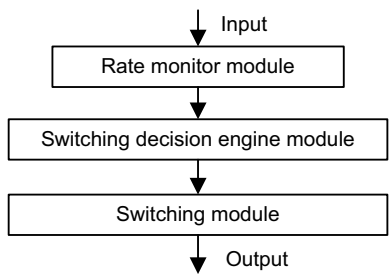

Figure 1. System block diagram

TABLE II. LIST OF NOTIONS

| Notation | Description |
|---|---|
| $T_{measure}$ | The data rate collect interval |
| $R_{user}$ | The data rate of the user |
| $B_t$ | The threshold of bandwidth that is set by the user |
| $P_{user}$ | The user's geographical location |
| $Lat_{user}$ | The latitude of the user's location |
| $Lon_{user}$ | The longitude of the user's location |
| $P_{ap}$ | The nearest Wi-Fi network AP's geographical location |
| $Lat_{ap}$ | The latitude of the nearest Wi-Fi network AP's location |
| $Lon_{ap}$ | The longitude of the nearest Wi-Fi network AP's location |
| $d$ | The distance between user and the nearest Wi-Fi network AP |
| $v_{user}$ | The speed of the user |
| $t_{switch}$ | The time needed to switch to Wi-Fi network |

On the mobile device, we periodically measure cellular data rates (both uplink and downlink) at a chosen interval $T_{measure}$ (see Table II).

Based on the collected information, the switching decision engine decides whether the users need switch to Wi-Fi network: if yes, then it invokes the switching model to find the nearest Wi-Fi network AP, and users switch to Wi-Fi network when they arrive at it.

*A. Rate Monitor Module*

To adapt to conditions change, our system must monitor the conditions of the mobile device and the wireless network. In [15] Yan et al. use available bandwidth, packet loss, received signal strength, energy consumption, operator requirements, user preferences, etc. to guide the selection of the best network. While in [16] Chamodrakas and Martakos propose a novel network selection based on TOPSIS [17] that takes into account both network conditions and user preferences as well as QoS and energy consumption to select the best network. However, in this paper, we simply use the data rate of the user to guide switching decision making. Whenever some significant change happens (for example, a large data rate of the user fluctuation occurs), the switching decision engine decides whether to trigger switching.

*B. Switching Decision Engine Module*

To perform switching decision engine, our system examines the data collected by rate monitor module. It then decides whether to trigger switching module according to user's switching goals. If so, it decides what level of $R_{user}$ to use on mobile device, that is, how much bandwidth the user needs. We use a simple threshold-based approach. If the $R_{user}$ is more than or equal to the threshold $B_t$, trigger switching module and switch to a wide bandwidth Wi-Fi network. To adapt to the user preferences, the user can set different thresholds and chooses the appropriate threshold.

*C. Switching Module*

In this module, our system decides when and how to switch to Wi-Fi networks. We present three different schemes which will be discussed in next section to switch from GPRS to Wi-Fi network. Fig. 2 shows the workflow of switching module of our proposed system, and it is discussed below.

When the switching module is triggered, it takes four steps to complete the switching process. In the first step, it uses A-GPS to get the user's geographical location $P_{user}(Lat_{user}, Lon_{user})$. In the second step, it finds the nearest Wi-Fi network AP by matching user's location $P_{user}$ in its log $L$ and returns the nearest Wi-Fi network AP's geographical position $P_{ap}(Lat_{ap}, Lon_{ap})$.

In the third step, it use (1) to calculate the distance between user and the nearest Wi-Fi network AP. In (1), $Lat_{user}$, $Lon_{user}$, $Lat_{ap}$ and $Lon_{ap}$ are expressed in radians. The unit of $d$ is kilometer.

$$d = 2\arcsin\sqrt{\sin^2\frac{Lat_{user}-Lat_{ap}}{2} + \cos(Lat_{user})\times\cos(Lat_{ap})\times\sin^2\frac{Lon_{user}-Lon_{ap}}{2}} \times 6378.137 \quad (1)$$

In the forth step, we simply assume that user moves towards the nearest Wi-Fi network AP at a constant speed $v_{user}$ and there are no obstacles between user and the nearest Wi-Fi network AP. It then uses the (2) to calculate the time needed to switch to Wi-Fi network. After the time $t_{switch}$, the user scans and connects to the Wi-Fi network.

$$t_{switch} = \frac{d}{v_{user}} \quad (2)$$

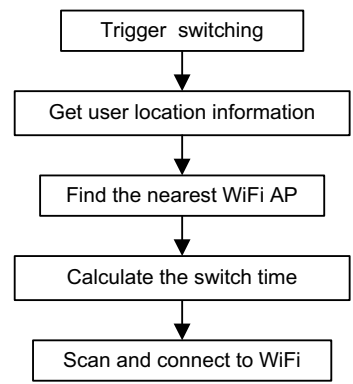

Figure 2. Workflow of switching module

## V. NETWORK SWITCHING SCHEME

In this section, we describe three switching schemes and two non-switching schemes. We will compare and discuss the power consumption of five schemes in next section.

### A. Switching Schemes

Here three switching schemes are considered. In the first scheme, denoted as *A-GPS assisted network switching scheme* (*A-GPS switching* for short), we use A-GPS to locate the user. Then, we find the nearest Wi-Fi network AP in the log by using the user's location information. Therefore, we can calculate the distance between user and nearest Wi-Fi network AP, and the time needed to switch to Wi-Fi network.

While in the second scheme, called *GSM assisted network switching scheme* (*GSM switching* for short), we use GSM to get the user's location.

In the third scheme, called *scanning assisted network switching scheme* (*Scanning switching* for short), we do not use the user's location information. When the switching module is triggered, the user scans for Wi-Fi networks. He or she connects to Wi-Fi network when he or she finds one. We simply assume that the user moves towards the nearest Wi-Fi network AP–that is, we measure the lower bound of power consumption.

### B. Non-switching Schemes

In this section, we describe two non-switching schemes as a comparison to three switching schemes. In the first scheme, called *always use GPRS* (*GPRS non-switching* for short), we don't switch and always use GPRS to access the internet. While in second scheme, called *always use Wi-Fi* (*Wi-Fi non-switching* for short), we always use Wi-Fi to surf the internet and also don't switch.

In order to compare the five schemes, we divide users into four categories according to their behaviors (see Table III). We will evaluate the five schemes in four thresholds (see Table IV) for different users in next section.

TABLE III. VARIOUS USER CHARACTERISTICS

| User | Applications | Bandwidth |
|---|---|---|
| U1 | Text message | Low |
| U2 | Text message and Web browsing | Medium |
| U3 | Text message and Video streaming | High |
| U4 | Text message and File download | Very High |

TABLE IV. DIFFERENT THRESHOLDS

| Threshold | Value |
|---|---|
| $B_1$ | 5 kb/s |
| $B_2$ | 10 kb/s |
| $B_3$ | 15 kb/s |
| $B_4$ | 20 kb/s |

## VI. PERFORMANCE EVALUATION

In this section, we use the log to evaluate our schemes for appropriate threshold and energy efficiency for different users.

### A. Experimental Setup

The experiment was based on a scenario that a user living in a suburban school area, as shown in Fig. 3, where each circle represents the approximate coverage area of each network.

We use ZTE X876 smartphone for our data collection. It is a Google Android 2.2 GSM phone with integrated Wi-Fi interface and is capable of EDGE data connectivity. It has a battery capacity of 1500mAh at 3.7 volts. We developed two logging softwares to record our experimental data with minimal intrusion to the normal smartphone operation.

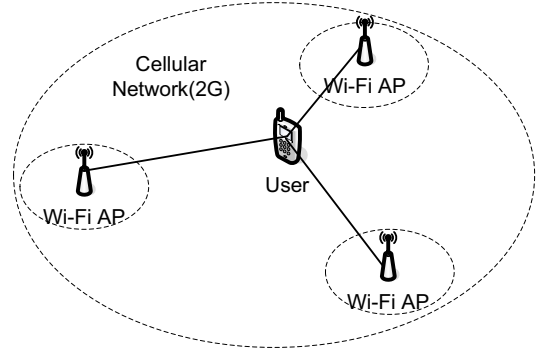

Figure 3. Networks used in experiments

*1) Rate Logger: Measuring Date Rates.* Our first logging software, called Rate Logger, measures different users' data rates every 30 seconds. When the experiment starts, Rate Logger begins to record user's data rate periodically. The first version of Rate Logger records user's data rate every 10 seconds. It is too frequently to represent the user's average data rate.

*2) Power Logger: Measuring Power Consumption.* We have developed software to measure the power consumption of our smartphone under controlled conditions. We use battery interface provided by Google APIs to record the current percent of the whole battery every minute.

### B. Energy Consumption

We evaluated the energy consumption of our scheme and compared it to other schemes for different users under different thresholds. Using the full battery capacity of the smartphone, we measure the power consumption of different schemes before the phone runs out of power (total capacity of 19980 J).

Fig. 4 shows the energy efficiency of five schemes for the user U1 under different thresholds $B_1$, $B_2$, $B_3$ and $B_4$. We observe that our scheme is close to other two switching schemes. Ideally, they are the same in terms of energy efficiency due to the user U1 does not need switch from GPRS to Wi-Fi network. Our scheme consumes a little more power than the best scheme of five schemes because of the overhead of our system itself. As we can see from Fig. 4, the user U1 does not need high bandwidth. Thus the bandwidth of GPRS is enough for U1, and the user U1 does not need any threshold.

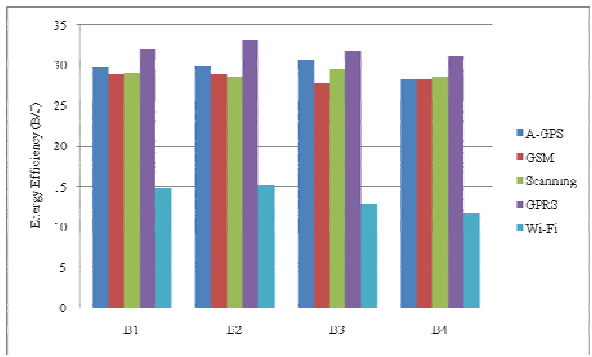

Figure 4. Energy efficiency of U1 under different thresholds

We plotted the energy efficiency of five schemes for the user U2 under different thresholds $B_1$, $B_2$, $B_3$ and $B_4$ in Fig. 5. As we can see from figure, our scheme is the best among the three switching schemes under the threshold $B_1$. There are two reasons behind this. First, A-GPS provides arguably the best combination of energy and accuracy for location sensing compare with GPS and GSM. Second, our scheme greatly reduces the numbers of unnecessary Wi-Fi scans on the mobile devices compare with other switching schemes. We observe that the user U2 doesn't need switch to Wi-Fi network under the thresholds $B_2$, $B_3$ and $B_4$. Therefore, the three switching schemes are close and the appropriate threshold for the user U2 is the threshold $B_1$.

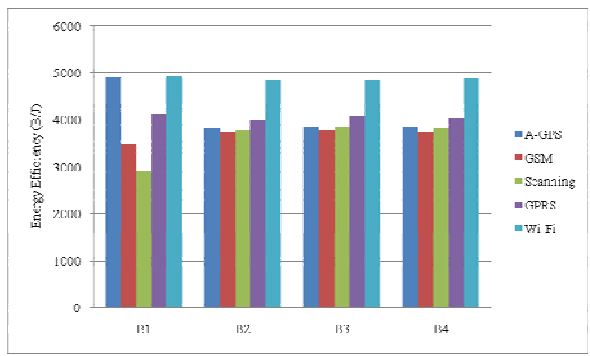

Figure 5. Energy efficiency of U2 under different thresholds

As shown in Fig. 6, our scheme is always the best among the three switching schemes. The reasons behind this were discussed before. As we can see from figure, the user U3 always switches from GPRS to Wi-Fi network under the different thresholds. Our scheme also consumes a little more energy than the best scheme of five schemes due to the overhead of our system itself and A-GPS. Fig. 6 shows the user U3 needs high bandwidth and the appropriate threshold for the user U3 is the threshold $B_4$.

The user U4 downloads a 50MB file from Internet. Fig. 7 shows the energy efficiency of five schemes for the user U4 under different thresholds $B_1$, $B_2$, $B_3$ and $B_4$. In our experiments, our scheme is always the best of three switching schemes. Because of many unnecessary Wi-Fi scans of the *scanning switching scheme*, it wastes too much energy. Due to the inaccuracy of GSM, the *GSM switching scheme* also does many unnecessary Wi-Fi scans on the mobile devices. This leads to its bad performance. Our scheme also consumes a little more energy than the best scheme–the Wi-Fi non-switching scheme. The reasons which cause this were discussed before. As we can see from figure, the user U4 needs very high bandwidth and the appropriate threshold for the user U4 is also the threshold $B_4$.

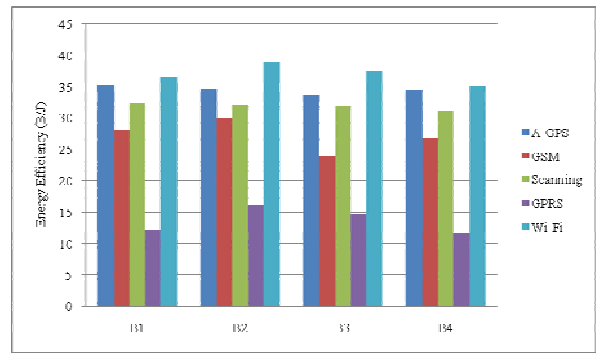

Figure 6. Energy efficiency of U3 under different thresholds

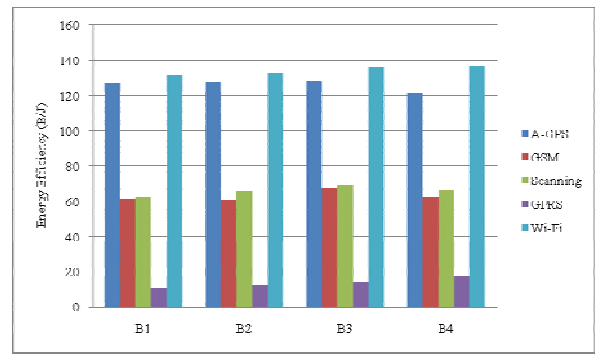

Figure 7. Energy efficiency of U4 under different thresholds

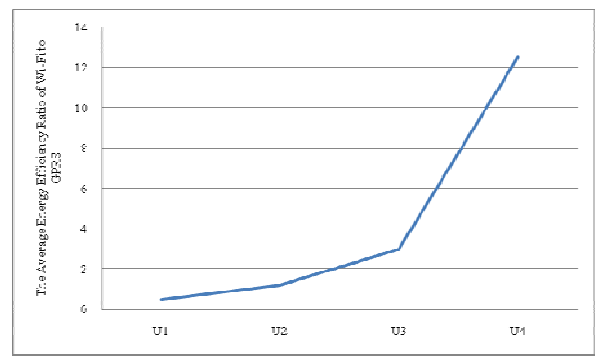

Figure 8. Average energy efficiency ratio of Wi-Fi to GPRS

As shown in Fig. 8, the GPRS has a better performance than the Wi-Fi network under the user U1. However, under the user U2, U3 and U4, the Wi-Fi network has a highly better performance than the GPRS. We observe that an approximate exponential increase in the average energy efficiency ratio of Wi-Fi to GPRS. Thus, we conclude that Wi-Fi provides the best combination of bandwidth and power efficiency for data transfers.

VII. DISCUSSION

In this section, we discuss extensions to our system along with future work. Our results show that our proposed scheme is always the best among the three switching schemes. However, in this paper, we simply use the data rate of the user to make switching decision. We don't take into account both network conditions and QoS to select the best network. Therefore, the switching decision making procedure must be intelligent enough to handle different conditions. Further studies of switching decision making algorithms are needed to address this challenge.

As we can see from our experiment, different users have different appropriate thresholds. Thus, our system must be adapted for different users. However, it is not intelligent enough to handle this. Further research is necessary to adapt the switching decision making algorithms for different users. Finally, we will upload every mobile device's log $L$ and offload the matching computations to the cloud computing platform in the future work.

VIII. CONCLUSION

In this paper, we have proposed an A-GPS assisted network switching scheme which uses A-GPS to find the nearest Wi-Fi network AP by using user's location information. We leverage complementary characteristics of Wi-Fi and GPRS to meet the user bandwidth needs and minimize the power consumption. Our proposed scheme allows the user to intelligently switch to the Wi-Fi interface when he or she arrives at the nearest Wi-Fi network AP. Thus, it greatly reduces the numbers of unnecessary Wi-Fi scans on non-connected state. We have implemented our scheme and our experiments show that it effectively saves energy for mobile devices. Finally, we also find different appropriate thresholds for different users.